\documentclass[aip,pof,12pt]{revtex4-1}
\usepackage{graphicx}
\usepackage{amssymb,amsmath}
\usepackage{natbib}
\usepackage{subfigure}
\usepackage{epstopdf}
\usepackage{color}

\newcommand{\nom}{Nu_\omega}
\newcommand{\usro}{Ro^{-1}}
\newcommand{\bu}{\boldsymbol{u}}

\begin{document}

\title{Turbulence decay towards the linearly-stable regime of Taylor-Couette flow}
\author{Rodolfo Ostilla M\'onico$^1$, Roberto Verzicco$^{1,2}$, Siegfried Grossmann$^{3}$ and Detlef Lohse$^1$}
\affiliation{
$^{1}$ Physics of Fluids Group, Mesa+ Institute,  and J.\ M.\ Burgers Centre for
Fluid Dynamics, University of Twente, 7500 AE Enschede, The Netherlands \\
$^2$ Dipartimento di Ingegneria Industriale, University of Rome ``Tor Vergata'', Via del Politecnico 1, Roma 00133, Italy \\
$^3$ Fachbereich Physik, Philipps-Universit\"{a}t Marburg, Am Renthof 6, D-35032 Marburg, Germany}
\date{\today}

\begin{abstract}
Taylor-Couette (TC) flow is used to probe the hydrodynamical stability of astrophysical
accretion disks. Experimental data on the subcritical stability of TC are in conflict 
about  the existence of turbulence (cf. Ji et al. 
Nature, \textbf{444}, 343-346 (2006) and Paoletti et al., A\&A, \textbf{547}, A64 (2012)),
with discrepancies attributed to end-plate effects.
In this paper we numerically simulate TC flow with axially periodic boundary conditions to explore
the existence of sub-critical transitions to turbulence when no end-plates are present. 
We start the simulations with a fully turbulent state in the unstable regime and enter the 
linearly stable regime by suddenly starting a (stabilizing) outer cylinder rotation.
The shear Reynolds number of the turbulent initial state is up to $Re_s \sim10^5$ and the radius ratio is $\eta=0.714$.
The stabilization causes the system to behave as a damped oscillator and correspondingly the turbulence 
decays. The evolution of the torque and turbulent kinetic 
energy is analysed and the periodicity and damping of the oscillations are quantified and explained
as a function of shear Reynolds number.
Though the initially turbulent flow state decays, surprisingly, the system is found to absorb energy during this decay.
\end{abstract}

\maketitle

Quasars are the most luminous objects in the Universe. They are thought to consist of supermassive
black holes in the centres of galaxies, which accrete matter and emit radiation, thereby transforming mass into energy
with an efficiency between 5-40\% \cite{ji13,arm11}. In order for orbiting material to fall into the so-called astrophysical accretion disk, 
it must lose its angular momentum. Molecular viscosity alone is not  enough to account for this loss,
so some sort of turbulent viscosity, causing the enhanced transport of angular momentum, has been 
conjectured \cite{sha73}, as otherwise gravitationally bound objects, like galaxies, stars or planets, would not exist. 
This implies that the flow of the material must be turbulent, but the origin of turbulence in accretion disks is currently
unknown. Accretion disks can either be composed of ``hot'' matter, which is fully ionized and electrically conducting, 
or ``cold'' matter, i.\ e., dust with less than 1\% ionized matter. Hot disks are found around quasars
and active galactic nuclei, and thought to undergo a magneto-rotational
instability (MRI) to become turbulent \cite{vel59,bal91b}. In contrast, cold disks are found in protoplanetary
systems, and are thought to have ``MRI-dead'' regions, where turbulence due the MRI cannot exist \cite{gam96,arm11}. 
The places where planets eventually form and reside coincide with these MRI-dead regions, 
so additional transport of angular momentum must take place. To account for this transport, another mechanism has been proposed:
hydrodynamical (HD) non-linear instabilities, i.e. transport by turbulence. 

Taylor-Couette flow (TC) is the flow between two coaxial cylinders which rotate independently.
It is used as a model for probing the HD stability of accretion disks \cite{ji13}.
Accretion disks have velocity profiles that are linearly stable,
but just as in  pipe and channel flows, the shear Reynolds numbers are so large that non-linear
instabilities may play a role in the formation of turbulence.
The question whether TC flow indeed does undergo  a non-linear transition 
\cite{tre93,gro00rmp} 
to turbulence 
at large shear 
Reynolds numbers has been probed experimentally by several authors with conflicting results\cite{ric01,dub05,ji06,gil11,gil12,pao11,pao12,avi12}. 
The discrepancies between the different experiments have been attributed to 
the end-plates of the TC devices, i.e. the solid boundaries which axially confine TC flow. These
cause secondary flows, known as Ekman circulation, which propagate into the flow and influence global stability.
Avila \cite{avi12}, based on his numerical simulations with end-plates imitating the Ji et al. \cite{ji06} and the Paoletti et al. \cite{pao12} experiments,
concluded that both of these experiments were not adequate for studying the non-linear transition due to the presence of end-plates. 
However, these simulations are at shear Reynolds numbers of the order $Re_s\sim 10^4$, while Ji et al. \cite{ji06} mention the counter-intuitive
result that $Re_s$ must \emph{exceed} a certain threshold before the flow relaminarizes.

Indeed, confinement of the flow plays a very important role in this transition, but it is unavoidable in experiments. Direct numerical 
simulations (DNS) do not have these limitations, as they can be run with periodic boundary conditions to completely prevent
end-plate effects. Following this idea, Lesur and Longaretti \cite{les05} simulated rotating 
plane Couette flow with perturbed initial conditions, going up to shear Reynolds numbers 
of order $10^4$, but due to computational limitations, the quasi-Keplerian regime velocity (i.e., the regime where the
two velocity boundary conditions are related by Kepler's third law), was not reached.

In this letter, we adopt a similar approach, but now for Taylor-Couette flow and in particular reaching
larger shear Reynolds numbers $Re_s \sim10^5$. We proceed as follows: We start from an already turbulent 
flow field and then add stabilizing outer cylinder rotation, wondering whether the turbulence is 
sustained. We find that it is not: the 
turbulence decays, and the torque decreases down to a value corresponding to purely azimuthal, laminar flow.
This means that not only the TC system is linearly stable in this geometry, but an initially turbulent flow even
\emph{decays} towards the linearly stable regime.

The DNSs were performed using a second-order finite difference code with fractional time-stepping \cite{ver96}. 
This code was already validated and used extensively in the context of TC \cite{ost13,ost14}. 
The turbulent initial conditions will be taken from previous work \cite{ost14}. The radius ratio
is $\eta=r_i/r_o=0.714$, where $r_i$ and $r_o$ are the inner and outer radius, respectively,
and the aspect ratio is $\Gamma=L/d=2.094$, where $L$ is the axial periodicity length, and $d$ the gap width $d=r_o-r_i$. A
rotational symmetry of order 6 was forced on the system to reduce computational costs while not affecting
the results \cite{bra13,ost14}.

The simulations are performed in a frame of reference co-rotating with the outer cylinder.
In the rotating frame, the inner cylinder has an
azimuthal velocity $U=r_i(\omega_i-\omega_o)$, with $\omega_i$ and $\omega_o$
the inner and outer cylinder angular velocities respectively, while the outer cylinder is now stationary. The resulting shear drives the flow,
and can be expressed non-dimensionally as a shear Reynolds number 
{\footnote{In the astrophysical context, e.g.\ ref.\  \cite{dub05}, often an extra factor $2/(1+\eta)$ is used in this 
definition.}} 
$Re_s=dU/\nu$, 
where $\nu$ is the
kinematic viscosity of the fluid.
In this rotating frame, the outer cylinder rotation in the lab frame manifests itself as a Coriolis force 
$Ro^{-1}(\bf{e_z} \times \bf{u})$,
where the Rossby number is defined as $Ro = U/(2\omega_o d)=(\eta[1-\mu])/(2\mu[1-\eta])$, 
with $\bf{e_z}$ the unit vector in the axial direction, and $\mu=\omega_o/\omega_i$ the angular 
velocity ratio (in the resting lab-system).
Note that $Ro$ can be either negative or positive. We define the non-dimensional radius as $\tilde{r}=(r-r_i)/d$, the non-dimensional
axial height as $\tilde{z}=z/d$ and the non-dimensional time as $\tilde{t}=tU/d$, where $d/U$ is the
large-eddy turnover time of the initial state.

Six simulations were run: Three with fixed 
$Re_s = 8.10\cdot10^3$,  but different strength of the Coriolis force, namely: 
a) $\usro=0.83$, corresponding to a system 
on the Rayleigh-stability line, equivalent to $r_i^2\omega_i = r_o^2\omega_o$, or $\mu=\eta^2\approx0.51$ in 
the lab frame of reference; 
b) $\usro=1.22$, corresponding to $\mu=\eta^{3/2}\approx0.60$
or a system in the quasi-Keplerian regime, 
i.e. the regime in which the angular velocities at the two cylinders are related by Kepler's third law
 $r_i^3\omega_o^2=r_o^3\omega_i^2$; 
 and 
c) $\usro=2.50$, an even stronger stabilization,
equivalent to $\mu\approx0.76$. The other three simulations were done by fixing $\usro=1.22$ and 
giving $Re_s$ the values 
$Re_s=8.10\cdot10^3$, 
$Re_s=2.52\cdot10^4$, and 
$Re_s=8.10\cdot10^4$.
Grids of $N_\theta\times N_r\times N_z$ = $256\times 640\times 512$ were used for all simulations, except for 
$Re_s=8.10\cdot10^4$, where
a grid of $N_\theta\times N_r\times N_z$ = $512\times 800\times 1024$ is required to guarantee sufficient resolution.

Simulating in a rotating frame adds a new perspective to the problem. In this frame the (non-dimensional) Navier-Stokes equations read: 

\begin{equation}
 \displaystyle
 \frac{\partial \hat{\bu}}{\partial \hat{t}} + \hat{\bu}\cdot\hat{\nabla}\hat{\bu} = -\hat{\nabla} \hat{p} +  
\displaystyle\frac{1}{Re_s} \hat{\nabla}^2\hat{\bu} + Ro^{-1} {\boldsymbol{e}_z}\times\hat{\bu}~.
\label{eq:rotatingTC}
\end{equation}
Strong enough outer cylinder co-rotation in the lab-frame implies  a large stabilizing Coriolis force $Ro^{-1}$ 
in the rotating frame 
and turbulence is suppressed \cite{ost13}. 
This argument for stabilization \cite{ost13} can be compared to that of Taylor-Proudman's theorem \cite{tay17,pro16}
which when applied to rotating Rayleigh-B\'enard flow implies that the flow is stable even at 
large thermal driving, as long as the background rotation is large enough\cite{cha81}.

\begin{figure}[ht]
  \centering
  \includegraphics[height=3.5cm]{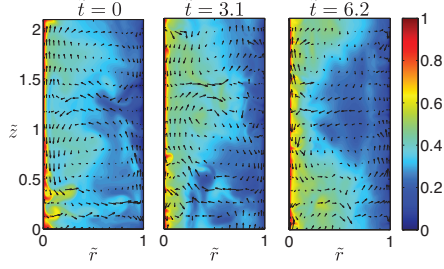}
  \caption{(color online)
Three contour plots of the instantaneous angular velocity field $\omega$ for 
$Re_s=8.10\cdot10^3$,
$\usro=1.22$, separated by $3.1$ nondimensional time units. Arrows represent the underlying axial and radial
velocities. The first snapshot shows the initial field, and in the other
two snapshots (at $\tilde{t}=3.1$ and $\tilde{t}=6.2$)
reversals of the Taylor rolls can be seen: The radial velocity in the gap 
at mid-height ($\tilde{z}\approx1.1$) changes from inwards
in the first snapshot, to outwards in the second, and back again to inwards in the last.
}
\label{fig:instomega}
\end{figure}

Fig.\ \ref{fig:instomega} shows three instantaneous snapshots of the angular velocity $\omega$ 
for 
$Re_s=8.10\cdot10^3$, one for $\tilde{t}=0$, and two for $\tilde{t}=3.1$ and $\tilde{t}=6.2$,
 after the stabilizing Coriolis force corresponding to the quasi-Keplerian regime, $\usro=1.22$,
was added. The sequence of figures shows the reversal of the Taylor vortices, which happens with a period of
$\tilde{t}\approx 6.2$. If the system is simulated for a large enough 
time, here $\tilde{t}\sim200$, the Taylor vortices  progressively fade away. The system
behaves as a damped oscillator.

To quantify these reversals, we define the quasi-Nusselt number \cite{eck07b} 
$\nom(r,t)$ as 
$\nom(r,t)=r^3(\langle u_r\omega\rangle_{\theta,z} - \nu \partial_r \langle \omega \rangle_{\theta,z})/T_{pa}$, 
where $T_{pa}$ is the torque required to drive the system in the purely azimuthal and laminar case. $\nom(r,z)$ represents
the transport of angular velocity. For a purely azimuthal flow with no turbulence,
$\nom=1$, so $\nom-1$ represents the additional transport of angular velocity due to turbulence. 

\begin{figure}[ht]
  \centering
  \includegraphics[width=0.35\textwidth]{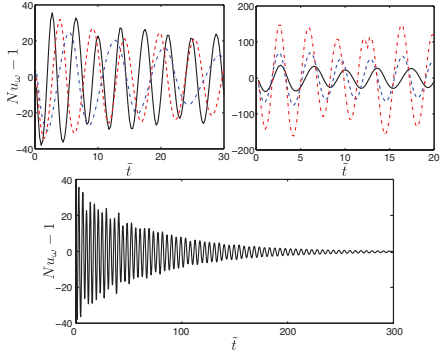}
  \caption{ (a): Time series of $\nom-1$ at mid-gap for 
$Re_s=8.10\cdot10^3$ and three values 
of $\usro$, $\usro=0.83$, (blue dash-dot line, Rayleigh-stability line), $\usro=1.22$, 
(quasi-Keplerian, i.e. stable, black line) and $\usro=2.50$ (even more stable, red dashed line), 
(b): Time series of $\nom(t)-1$  for $\usro=1.22$ and three values of $Re_s$
(black line: 
$Re_s=8.10\cdot10^3$, blue dashed line: $Re_s=2.52\cdot10^4$, and red dash-dot line: $Re_s=8.10\cdot10^4$),
(c): longer time series of (a) for $\usro=1.22$. 
On all figures (a), (b), (c) large oscillations of $\nom-1$ from positive to negative can be seen, which, when averaged, show that there is no 
net turbulent transport of angular velocity. For large enough time, the fluctuations decay and $\nom-1$ drops to zero.}
\label{fig:instnura}
\end{figure}

Figure \ref{fig:instnura}a
shows a time series of the axially averaged $\nom-1$ at the mid-radius $\tilde{r}=0.5$,
for 
$Re_s=8.10\cdot10^3$ and three values of the Coriolis force. $\nom-1$ can be seen to oscillate between large positive 
and large negative values, with an average 
 $\approx 0$, i.e. no net turbulent transport. The period of oscillation depends on $\usro$. Fig.\ \ref{fig:instnura}c shows long time
series for $\usro=1.22$. For large enough times, the oscillations damp out and the flow becomes purely azimuthal and
laminar, $\nom=1$. Fig.\ \ref{fig:instnura}b
 shows $\nom-1$ for three values of $Re_s$ and $\usro=1.22$. Similar oscillations
in $\nom-1$ can be seen, but in this case the period weakly depends on $Re_s$.

\begin{figure}[ht]
  \centering
  \includegraphics[width=0.47\textwidth]{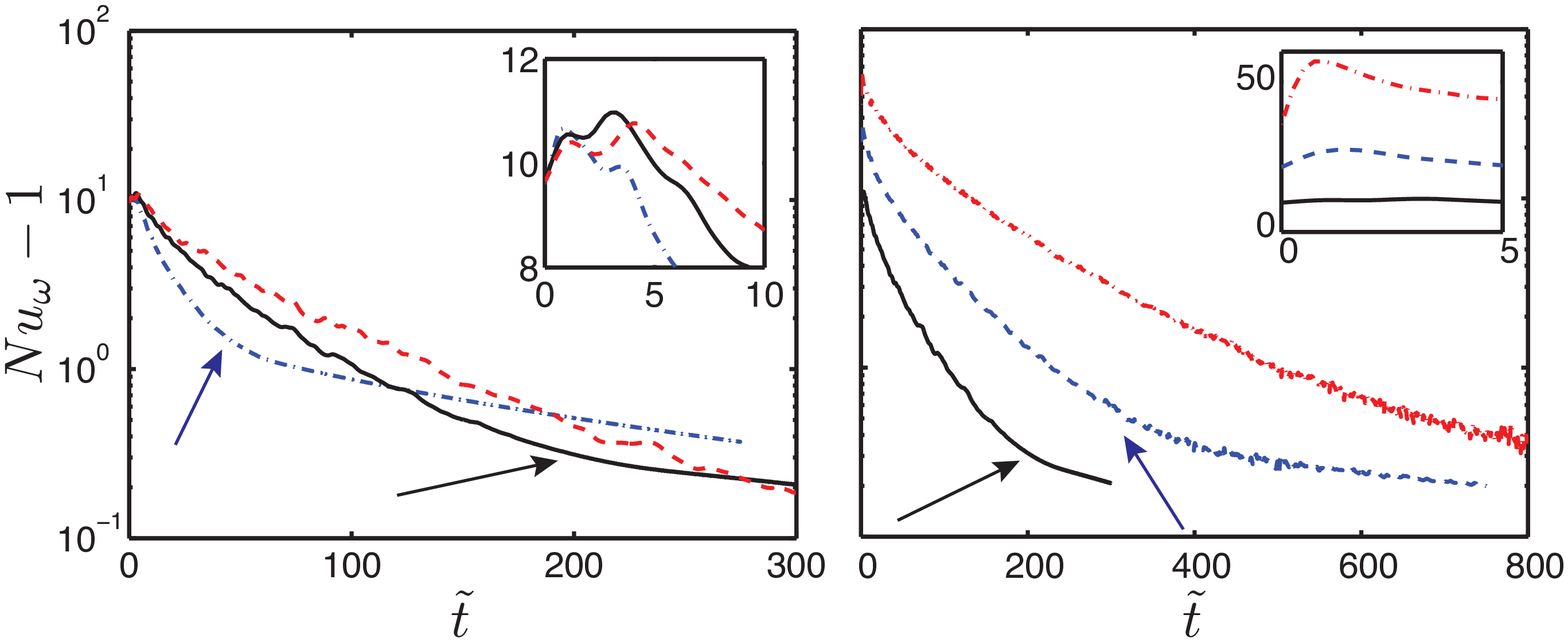}
  \caption{ (a): semi-log time series of $\nom-1$ at 
the inner cylinder ($\tilde{r}=0$, $z$-averaged) for 
$Re_s=8.10\cdot10^3$ and three values 
of $\usro$, namely $\usro=0.83$, (Rayleigh-stability line, blue dash-dot line), $\usro=1.22$
(quasi-Keplerian, black line), and $\usro=2.50$ (red dashed line). (b): semi-log time series of $\nom-1$ at the inner cylinder for $\usro=1.22$
and three values of 
$Re_s$ (black line: $Re_s=8.10\cdot10^3$, blue dashed line: $Re_s=2.52\cdot10^4$, and 
red dash-dot line: $Re_s=8.10\cdot10^4$).
Two decay modes can be seen in this figure, with the crossovers marked with  arrows. 
In both (a), (b), fluctuations are absent, unlike 
at the mid-gap, but $Nu_\omega - 1$  decays. For large enough time, $\nom-1$ drops to zero, taking
more time to do so for higher $Re_s$ or smaller $\usro$. Zoom-ins of the first few large eddy turnover times are shown 
as insets (linear scale) for both figures.
Transients, corresponding to an increase of $\nom$ can be seen during the initial stages.}
\label{fig:instnuic}
\end{figure}

Figure \ref{fig:instnuic} shows the axially averaged $\nom-1$ but now measured directly at the inner cylinder (probing
the inner boundary layer (BL)), for different
values of $\usro$ and $Re_s$. Unlike the previous mid gap case, no oscillations 
can be seen, reflecting that the oscillations only occur in the bulk of the flow, where the Taylor vortices are,
but that the BLs are unaffected. Of course, the  decay towards the purely azimuthal, laminar state $\nom=1$ is also 
observed in the BLs (Fig \ref{fig:instnuic}a,b), independent of $\usro$ and $Re_s$. 
 
\begin{figure}[ht]
  \centering
  \includegraphics[width=0.47\textwidth]{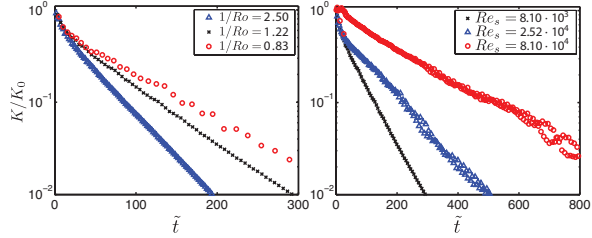}
  \caption{ (a): Time series of $K(\tilde{t})/K_0$ for 
$Re_s=8.10\cdot10^3$ and three values 
of $\usro$, $\usro=0.83$, (Rayleigh-stability line), $\usro=1.22$, 
(quasi-Keplerian) and $\usro=2.50$. (b): Time series of $K/K_0$ for $\usro=1.22$
and three values of $Re_s$. After the initial transient, an exponential decay of the turbulent kinetic 
energy can be seen, even at the largest Reynolds numbers.}
\label{fig:uwind}
\end{figure}

Finally, we quantify the decay of turbulence. To do so, we define the turbulent kinetic energy substitute of the flow
as $K(t)=\frac{1}{2}(u_r^{2}+u_z^{2})_{r,z,\theta}$. Note that this is not the full turbulent energy, as for that
also fluctuations in $u_\theta$ contribute, but, by definition, $K$ has the nice property to be zero in the purely azimuthal
state. Fig.\ \ref{fig:uwind}a shows
the decay of $K$ for 
$Re_s=8.10\cdot10^3$ and three values of $\usro$, while fig \ref{fig:uwind}b shows the 
decay for $\usro=1.22$ and three values of $Re_s$. 
In the simulations, $K(\tilde{t})$ shows short time scale fluctuations, similar to that of $\nom$
in the mid-gap, so for clarity, only the peak of $K(\tilde{t})$ in every cycle is represented.

\begin{figure}[ht]
  \centering
  \includegraphics[width=0.47\textwidth]{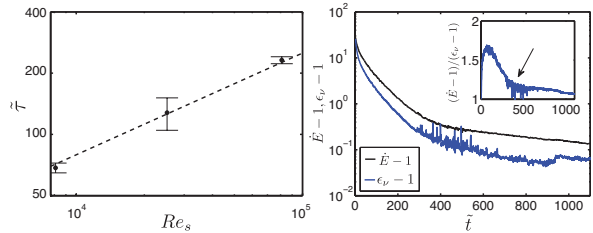}
   \caption{ 
(a) log-log plot of the turbulence decay time $\tilde{\tau}$ vs $Re_s$ for 
$\usro=1.22$ (quasi-Keplerian) and 
the three values of $Re_s$ simulated. 
The power law $\tilde{\tau} = 0.79\sqrt{Re_s}$ fits the data points. (b): Semi-logarithmic
plot of the time series of both the energy input through the boundaries $\dot{E}-1$ and the viscid dissipation of energy in the system $\tilde \epsilon_\nu-1$ for
$Re_s=2.49\cdot10^4$ and $\usro=1.22$. The laminar dissipation, i.e. $1$ with the non-dimensionalization chosen, is subtracted
from both variables. An inset with the ratio between both energies is shown. The arrow marks
the transition between the two decay mechanisms. For quite a long time, the system still absorbs more
energy than it dissipates; equilibrium is reached only for $\tilde{t}$ of order beyond $1000$, due to the small decay rate.  }
\label{fig:tau}
\end{figure}

As expected from the previous results on $\nom(\tilde{t})$, $K(\tilde{t})$ decays to zero when given enough time. 
The only mechanism for energy dissipation is viscosity, so it is not surprising that the decay time is proportional to 
the energy content, thus leading to an approximately exponential decay, $K(\tilde{t})=K_0\exp(-\tilde{t}/\tilde{\tau})$
where $\tilde{\tau}$ is the characteristic (dimensionless) decay time, and $K_0=K(\tilde{t}=0)$. 
Fits to the data in fig.\ \ref{fig:uwind}b were performed to obtain
estimates for the dependence $\tilde{\tau}(Re_s)$, which is shown in fig.\ \ref{fig:tau}a.
The absolute value of $\tilde{\tau}$ is large (around $70$ large eddy turnover 
times for $Re_s\sim10^4$ and $\usro=1.22$) and in the range studied 
shows a power-law  \begin{equation} \tilde{\tau}\sim\sqrt{Re_s}. \label{einhalb} \end{equation} 
This scaling can be understood from realizing that the decay of the turbulence is first dominated by the
energy dissipation rate $\epsilon_{BL}$ in the BLs, which scales as \cite{gro00} 
$\epsilon_{BL} \sim \nu^3 d^{-4} Re_s^{5/2}$. This immediately implies the scaling relation  (\ref{einhalb}) for the 
typical decay time $\tau = \tilde \tau d/U \sim U^2/\epsilon_{BL}$. Only for very large times or for very small
Reynolds numbers from the very beginning, i.e.,  
when the thickness of 
the BLs is comparable to the gap thickness $d$ and the Taylor rolls have died out or are absent, one has 
$\epsilon_{BL} \sim \nu^3 d^{-4} Re_s^2$ (cf.\ ref.\ \cite{gro01}) and thus $\tau\sim Re_s$. 
This second regime can be seen in figs.\ \ref{fig:instnuic}b and \ref{fig:tau}b.

We highlight that even though $K(\tilde{t})$ decays, the total energy of the system \emph{increases}, as it absorbs more energy
through the boundaries than it dissipates viscidly. This can be seen in fig.\  \ref{fig:tau}b,
displaying a time series of the non-dimensionalized energy input into the system $\dot{E} = (T_i\omega_i+T_o\omega_o) / (\rho\Omega\epsilon_{pa})$, where $T_{i,o}$ is the
torque applied at the inner (outer) cylinder, $\rho$ and $\Omega$ are the fluid density and volume, respectively, and
$\epsilon_{pa}$ is the viscous dissipation in the purely azimuthal and laminar state, and a time series of the non-dimensionalized energy dissipation 
rate inside the system,
$\tilde{\epsilon}_\nu=\frac{1}{2}\nu\langle[\partial_i u_j + \partial_j u_i]^2\rangle_\Omega/\epsilon_{pa}$.
When transitioning from turbulence
to the purely azimuthal $u_\theta(r)$-profile, the system \emph{absorbs} energy, while at the same time the initial turbulence \emph{decays}.
The purely azimuthal $u_\theta(r)$-profile has more energy than its turbulent counterpart. Indeed, this is
the reason for the Rayleigh instability, now supressed by the Coriolis force.

Again, two time scales for the decay can be seen in the inset of fig \ref{fig:tau}b. 
After the initial transient, turbulence in the bulk decays fast, due to the efficient angular momentum transport by the Taylor rolls. 
After about 200-300 large eddy turnover times 
this decay mechanism is exhausted as then the (turbulent) Taylor rolls in the bulk are so weak that they
can no longer transport angular momentum. Then
a second decay mechanism takes over, whose onset is marked by an arrow in the inset of fig \ref{fig:tau}b.
In this regime, characterized 
 by purely viscous dissipation  $\sim \nu^3 L^{-4} Re_s^2$ in the whole gap,  
the profile  $u_\theta(r)$ returns to the azimuthal and laminar one and 
$\nom$ thus to $1$.
It is important to note that $\nom$ \emph{must} return to $1$, as in the statistically stationary state, $\nom$ has to be independent of $r$.

In summary, consistent with what had been found in the experiments of Ji et al.\cite{ji06}, no turbulence, and therefore, no turbulent
transport of angular momentum can be seen in the Rayleigh-stable regime for the control parameter range studied once the effect of axial boundaries is mitigated.
An initially turbulent state is stabilized by the corotation of the outer cylinder and the
transport of angular momentum in the bulk ceases; the turbulence decays in all our simulations, 
for $Re_s$ up to $8.10\cdot 10^4$. 

We point out, however, that even though the system is stable, the decay times for turbulence are long,
ranging between hundreds and thousands of large eddy turnover times. If one extrapolates 
{\footnote{
I.e., ignoring possible crossovers to an ultimate regime in which the energy dissipation rate may be dominated 
by the bulk dissipation \cite{gro00,gro11} $\epsilon_{bulk}\sim  \nu^3 d^{-4} Re_s^3$. In that regime $\tilde\tau$ would become
independent on $Re_s$. 
}}
the scaling 
relation (\ref{einhalb}) 
to shear Reynolds numbers attainable in cold accretion disks, which are of the order $10^{14}$, decay times
for the turbulence become very large, and the system might not have time to stabilize before enough angular momentum is transported
and the disk collapses into a planetary system.  It is also unclear whether the extreme values of $\eta$ present in astrophysical
accretion disks will have an effect on the decay times. The Rossby number has a strong $\eta$-dependence, and for $\eta\to 0$ it diverges.
Also, other types of mechanisms might dominate the angular momentum transport in cold accretion disks, 
like subcritical baroclinic instabilities which lead to turbulence \cite{wit74,kla03} or transport associated
to self-gravity \cite{pac81}. We refer the reader to the review by Armitage \cite{arm11} for a more detailed discussion.

In conclusion, in our TC simulations without end-plates, in the parameter range studied up to shear Reynolds 
numbers of $10^5$, the flow was seen to not only remain 
laminar in the presence of small perturbations, but when starting from an initially turbulent state 
we can see that the turbulence decays and the system returns to its laminar state. All of this happens while the system absorbs energy. 
Therefore, we can attribute the turbulence and increased
angular momentum transport found in some of the experiments \cite{dub05,pao11} and numerical simulations \cite{avi12}
which probe the linearly stable regime of TC, to  effects of the end-plate confinement.

Acknowledgments: We would like to thank C. Sun for various stimulating discussions.
We acknowledge an ERC Advanced Grant and the PRACE resource 
Curie, based in France at GENCI/CEA.

\bibliography{literatur}

\begin{thebibliography}{10}%
\makeatletter
\providecommand \@ifxundefined [1]{%
 \ifx #1\undefined \expandafter \@firstoftwo
 \else \expandafter \@secondoftwo
\fi
}%
\providecommand \@ifnum [1]{%
 \ifnum #1\expandafter \@firstoftwo
 \else \expandafter \@secondoftwo
\fi
}%
\providecommand \enquote [1]{``#1''}%
\providecommand \bibnamefont  [1]{#1}%
\providecommand \bibfnamefont [1]{#1}%
\providecommand \citenamefont [1]{#1}%
\providecommand\href[0]{\@sanitize\@href}%
\providecommand\@href[1]{\endgroup\@@startlink{#1}\endgroup\@@href}%
\providecommand\@@href[1]{#1\@@endlink}%
\providecommand \@sanitize [0]{\begingroup\catcode`\&12\catcode`\#12\relax}%
\@ifxundefined \pdfoutput {\@firstoftwo}{%
 \@ifnum{\z@=\pdfoutput}{\@firstoftwo}{\@secondoftwo}%
}{%
 \providecommand\@@startlink[1]{\leavevmode\special{html:<a href="#1">}}%
 \providecommand\@@endlink[0]{\special{html:</a>}}%
}{%
 \providecommand\@@startlink[1]{%
  \leavevmode
  \pdfstartlink
   attr{/Border[0 0 1 ]/H/I/C[0 1 1]}%
   user{/Subtype/Link/A<</Type/Action/S/URI/URI(#1)>>}%
  \relax
 }%
 \providecommand\@@endlink[0]{\pdfendlink}%
}%
\providecommand \url  [0]{\begingroup\@sanitize \@url }%
\providecommand \@url [1]{\endgroup\@href {#1}{\urlprefix}}%
\providecommand \urlprefix [0]{URL }%
\providecommand \Eprint[0]{\href }%
\@ifxundefined \urlstyle {%
  \providecommand \doi [1]{doi:\discretionary{}{}{}#1}%
}{%
  \providecommand \doi [0]{doi:\discretionary{}{}{}\begingroup
  \urlstyle{rm}\Url }%
}%
\providecommand \doibase [0]{http://dx.doi.org/}%
\providecommand \Doi[1]{\href{\doibase#1}}%
\providecommand \selectlanguage [0]{\@gobble}%
\providecommand \bibinfo [0]{\@secondoftwo}%
\providecommand \bibfield [0]{\@secondoftwo}%
\providecommand \translation [1]{[#1]}%
\providecommand \BibitemOpen[0]{}%
\providecommand \bibitemStop [0]{}%
\providecommand \bibitemNoStop [0]{.\EOS\space}%
\providecommand \EOS [0]{\spacefactor3000\relax}%
\providecommand \BibitemShut [1]{\csname bibitem#1\endcsname}%
\bibitem{ji13}%
  \BibitemOpen
  \bibfield{author}{%
  \bibinfo {author} {\bibfnamefont{H.}~\bibnamefont{Ji}}\ and\ \bibinfo
  {author} {\bibfnamefont{S.~A.}\ \bibnamefont{Balbus}},\ }%
  \bibfield{title}{%
  \enquote{\bibinfo {title} {Angular momentum transport in astrophysics and in
  the lab},}\ }%
  \bibfield{journal}{%
  \bibinfo {journal} {Phys. Today}\ }%
  \textbf{\bibinfo {volume} {66}},\ \bibinfo {pages} {27--33} (\bibinfo {month}
  {Aug}\ \bibinfo {year} {2013})\BibitemShut{NoStop}%
\bibitem{arm11}%
  \BibitemOpen
  \bibfield{author}{%
  \bibinfo {author} {\bibfnamefont{P.~J.}\ \bibnamefont{Armitage}},\ }%
  \bibfield{title}{%
  \enquote{\bibinfo {title} {Dynamics of protoplanetary disks},}\ }%
  \bibfield{journal}{%
  \bibinfo {journal} {Ann. Rev. Astron. Astrophys.}\ }%
  \textbf{\bibinfo {volume} {49}} (\bibinfo {year} {2011})\BibitemShut{NoStop}%
\bibitem{sha73}%
  \BibitemOpen
  \bibfield{author}{%
  \bibinfo {author} {\bibfnamefont{N.~I.}\ \bibnamefont{Shakura}}\ and\
  \bibinfo {author} {\bibfnamefont{R.~A.}\ \bibnamefont{Sunyaev}},\ }%
  \bibfield{title}{%
  \enquote{\bibinfo {title} {Black holes in binary systems. observational
  appearance.}.}\ }%
  \bibfield{journal}{%
  \bibinfo {journal} {Astron. \& Astrophys}\ }%
  \textbf{\bibinfo {volume} {24}},\ \bibinfo {pages} {337--355} (\bibinfo
  {year} {1973})\BibitemShut{NoStop}%
\bibitem{vel59}%
  \BibitemOpen
  \bibfield{author}{%
  \bibinfo {author} {\bibfnamefont{E.~P.}\ \bibnamefont{Velikhov}},\ }%
  \bibfield{title}{%
  \enquote{\bibinfo {title} {Stability of an ideally conducting liquid flowing
  between rotating cylinders in a magnetic field},}\ }%
  \bibfield{journal}{%
  \bibinfo {journal} {Zhur. Eksptl'. i Teoret. Fiz.}\ }%
  \textbf{\bibinfo {volume} {36}} (\bibinfo {year} {1959})\BibitemShut{NoStop}%
\bibitem{bal91b}%
  \BibitemOpen
  \bibfield{author}{%
  \bibinfo {author} {\bibfnamefont{S.~A.}\ \bibnamefont{Balbus}}\ and\ \bibinfo
  {author} {\bibfnamefont{J.}~\bibnamefont{Hawley}},\ }%
  \bibfield{title}{%
  \enquote{\bibinfo {title} {A powerful local shear instability in weakly
  magnetized disks},}\ }%
  \bibfield{journal}{%
  \bibinfo {journal} {Astrophys. J.}\ }%
  \textbf{\bibinfo {volume} {376}} (\bibinfo {year}
  {1991})\BibitemShut{NoStop}%
\bibitem{gam96}%
  \BibitemOpen
  \bibfield{author}{%
  \bibinfo {author} {\bibfnamefont{C.~F.}\ \bibnamefont{Gammie}},\ }%
  \bibfield{title}{%
  \enquote{\bibinfo {title} {Layered accretion in {{{T-Tauri}}} disks},}\ }%
  \bibfield{journal}{%
  \bibinfo {journal} {Astrophys. J.}\ }%
  \textbf{\bibinfo {volume} {457}} (\bibinfo {year}
  {1996})\BibitemShut{NoStop}%
\bibitem{tre93}%
  \BibitemOpen
  \bibfield{author}{%
  \bibinfo {author} {\bibfnamefont{L.~N.}\ \bibnamefont{Trefethen}}, \bibinfo
  {author} {\bibfnamefont{A.~E.}\ \bibnamefont{Trefethen}}, \bibinfo {author}
  {\bibfnamefont{S.~C.}\ \bibnamefont{Reddy}},\ and\ \bibinfo {author}
  {\bibfnamefont{T.~A.}\ \bibnamefont{Driscol}},\ }%
  \bibfield{title}{%
  \enquote{\bibinfo {title} {Hydrodynamic stability without eigenvalues},}\ }%
  \bibfield{journal}{%
  \bibinfo {journal} {Science}\ }%
  \textbf{\bibinfo {volume} {261}},\ \bibinfo {pages} {578--584} (\bibinfo
  {year} {1993})\BibitemShut{NoStop}%
\bibitem{gro00rmp}%
  \BibitemOpen
  \bibfield{author}{%
  \bibinfo {author} {\bibfnamefont{S.}~\bibnamefont{Grossmann}},\ }%
  \bibfield{title}{%
  \enquote{\bibinfo {title} {The onset of shear flow turbulence},}\ }%
  \bibfield{journal}{%
  \bibinfo {journal} {Rev. Mod. Phys.}\ }%
  \textbf{\bibinfo {volume} {72}},\ \bibinfo {pages} {603--618} (\bibinfo
  {year} {2000})\BibitemShut{NoStop}%
\bibitem{ric01}%
  \BibitemOpen
  \bibfield{author}{%
  \bibinfo {author} {\bibfnamefont{D.}~\bibnamefont{Richard}},\ }%
  \emph{\bibinfo {title} {Instabilit{\'e}s Hydrodynamiques dans les Ecoulements
  en Rotation Diff{\'e}rentielle}},\ Ph.D. thesis,\ \bibinfo {school} {Univ.
  Paris 7} (\bibinfo {year} {2001})\BibitemShut{NoStop}%
\bibitem{dub05}%
  \BibitemOpen
  \bibfield{author}{%
  \bibinfo {author} {\bibfnamefont{B.}~\bibnamefont{Dubrulle}}, \bibinfo
  {author} {\bibfnamefont{O.}~\bibnamefont{Dauchot}}, \bibinfo {author}
  {\bibfnamefont{F.}~\bibnamefont{Daviaud}}, \bibinfo {author}
  {\bibfnamefont{P.~Y.}\ \bibnamefont{Longaretti}}, \bibinfo {author}
  {\bibfnamefont{D.}~\bibnamefont{Richard}},\ and\ \bibinfo {author}
  {\bibfnamefont{J.~P.}\ \bibnamefont{Zahn}},\ }%
  \bibfield{title}{%
  \enquote{\bibinfo {title} {Stability and turbulent transport in
  {{Taylor--Couette}} ﬂow from analysis of experimental data},}\ }%
  \bibfield{journal}{%
  \bibinfo {journal} {Phys. Fluids}\ }%
  \textbf{\bibinfo {volume} {17}},\ \bibinfo {pages} {095103} (\bibinfo {year}
  {2005})\BibitemShut{NoStop}%
\bibitem{ji06}%
  \BibitemOpen
  \bibfield{author}{%
  \bibinfo {author} {\bibfnamefont{H.}~\bibnamefont{Ji}}, \bibinfo {author}
  {\bibfnamefont{M.}~\bibnamefont{Burin}}, \bibinfo {author}
  {\bibfnamefont{E.}~\bibnamefont{Schartman}},\ and\ \bibinfo {author}
  {\bibfnamefont{J.}~\bibnamefont{Goodman}},\ }%
  \bibfield{title}{%
  \enquote{\bibinfo {title} {{Hydrodynamic turbulence cannot transport angular
  momentum effectively in astrophysical disks}},}\ }%
  \bibfield{journal}{%
  \Doi{10.1038/nature05323}{\bibinfo {journal} {Nature}}\ }%
  \textbf{\bibinfo {volume} {444}},\ \bibinfo {pages} {343--346} (\bibinfo
  {year} {2006})\BibitemShut{NoStop}%
\bibitem{gil11}%
  \BibitemOpen
  \bibfield{author}{%
  \bibinfo {author} {\bibfnamefont{D.~P.~M.}\ \bibnamefont{van Gils}}, \bibinfo
  {author} {\bibfnamefont{S.~G.}\ \bibnamefont{Huisman}}, \bibinfo {author}
  {\bibfnamefont{G.~W.}\ \bibnamefont{Bruggert}}, \bibinfo {author}
  {\bibfnamefont{C.}~\bibnamefont{Sun}},\ and\ \bibinfo {author}
  {\bibfnamefont{D.}~\bibnamefont{Lohse}},\ }%
  \bibfield{title}{%
  \enquote{\bibinfo {title} {Torque scaling in turbulent {{Taylor-Couette}}
  flow with co- and counter-rotating cylinders},}\ }%
  \bibfield{journal}{%
  \bibinfo {journal} {Phys. Rev. Lett.}\ }%
  \textbf{\bibinfo {volume} {106}},\ \bibinfo {pages} {024502} (\bibinfo {year}
  {2011})\BibitemShut{NoStop}%
\bibitem{gil12}%
  \BibitemOpen
  \bibfield{author}{%
  \bibinfo {author} {\bibfnamefont{D.~P.~M.}\ \bibnamefont{van Gils}}, \bibinfo
  {author} {\bibfnamefont{S.~G.}\ \bibnamefont{Huisman}}, \bibinfo {author}
  {\bibfnamefont{S.}~\bibnamefont{Grossmann}}, \bibinfo {author}
  {\bibfnamefont{C.}~\bibnamefont{Sun}},\ and\ \bibinfo {author}
  {\bibfnamefont{D.}~\bibnamefont{Lohse}},\ }%
  \bibfield{title}{%
  \enquote{\bibinfo {title} {Optimal {{Taylor-Couette}} turbulence},}\ }%
  \bibfield{journal}{%
  \bibinfo {journal} {J. Fluid Mech.}\ }%
  \textbf{\bibinfo {volume} {706}},\ \bibinfo {pages} {118--149} (\bibinfo
  {year} {2012})\BibitemShut{NoStop}%
\bibitem{pao11}%
  \BibitemOpen
  \bibfield{author}{%
  \bibinfo {author} {\bibfnamefont{M.~S.}\ \bibnamefont{Paoletti}}\ and\
  \bibinfo {author} {\bibfnamefont{D.~P.}\ \bibnamefont{Lathrop}},\ }%
  \bibfield{title}{%
  \enquote{\bibinfo {title} {Angular momentum transport in turbulent flow
  between independently rotating cylinders},}\ }%
  \bibfield{journal}{%
  \bibinfo {journal} {Phys. Rev. Lett.}\ }%
  \textbf{\bibinfo {volume} {106}},\ \bibinfo {pages} {024501} (\bibinfo {year}
  {2011})\BibitemShut{NoStop}%
\bibitem{pao12}%
  \BibitemOpen
  \bibfield{author}{%
  \bibinfo {author} {\bibfnamefont{M.~S.}\ \bibnamefont{Paoletti}}, \bibinfo
  {author} {\bibfnamefont{D.~P.~M.}\ \bibnamefont{van Gils}}, \bibinfo {author}
  {\bibfnamefont{B.}~\bibnamefont{Dubrulle}}, \bibinfo {author}
  {\bibfnamefont{C.}~\bibnamefont{Sun}}, \bibinfo {author}
  {\bibfnamefont{D.}~\bibnamefont{Lohse}},\ and\ \bibinfo {author}
  {\bibfnamefont{D.~P.}\ \bibnamefont{Lathrop}},\ }%
  \bibfield{title}{%
  \enquote{\bibinfo {title} {Angular momentum transport and turbulence in
  laboratory models of keplerian flows},}\ }%
  \bibfield{journal}{%
  \bibinfo {journal} {Astron. \& Astrophys}\ }%
  \textbf{\bibinfo {volume} {547}},\ \bibinfo {pages} {A64} (\bibinfo {year}
  {2012})\BibitemShut{NoStop}%
\bibitem{avi12}%
  \BibitemOpen
  \bibfield{author}{%
  \bibinfo {author} {\bibfnamefont{M.}~\bibnamefont{Avila}},\ }%
  \bibfield{title}{%
  \enquote{\bibinfo {title} {Stability and angular-momentum transport of fluid
  flows between corotating cylinders},}\ }%
  \bibfield{journal}{%
  \bibinfo {journal} {Phys. Rev. Lett.}\ }%
  \textbf{\bibinfo {volume} {108}},\ \bibinfo {pages} {124501} (\bibinfo {year}
  {2013})\BibitemShut{NoStop}%
\bibitem{les05}%
  \BibitemOpen
  \bibfield{author}{%
  \bibinfo {author} {\bibfnamefont{G.}~\bibnamefont{Lesur}}\ and\ \bibinfo
  {author} {\bibfnamefont{P.~Y.}\ \bibnamefont{Longaretti}},\ }%
  \bibfield{title}{%
  \enquote{\bibinfo {title} {On the relevance of subcritical hydrodynamic
  turbulence to accretion disk transport},}\ }%
  \bibfield{journal}{%
  \bibinfo {journal} {Astron. \& Astrophys}\ }%
  \textbf{\bibinfo {volume} {444}},\ \bibinfo {pages} {25--44} (\bibinfo {year}
  {2005})\BibitemShut{NoStop}%
\bibitem{ver96}%
  \BibitemOpen
  \bibfield{author}{%
  \bibinfo {author} {\bibfnamefont{R.}~\bibnamefont{Verzicco}}\ and\ \bibinfo
  {author} {\bibfnamefont{P.}~\bibnamefont{Orlandi}},\ }%
  \bibfield{title}{%
  \enquote{\bibinfo {title} {A finite-difference scheme for three-dimensional
  incompressible flow in cylindrical coordinates},}\ }%
  \bibfield{journal}{%
  \bibinfo {journal} {J. Comput. Phys.}\ }%
  \textbf{\bibinfo {volume} {123}},\ \bibinfo {pages} {402--413} (\bibinfo
  {year} {1996})\BibitemShut{NoStop}%
\bibitem{ost13}%
  \BibitemOpen
  \bibfield{author}{%
  \bibinfo {author} {\bibfnamefont{R.}~\bibnamefont{Ostilla}}, \bibinfo
  {author} {\bibfnamefont{R.~J. A.~M.}\ \bibnamefont{Stevens}}, \bibinfo
  {author} {\bibfnamefont{S.}~\bibnamefont{Grossmann}}, \bibinfo {author}
  {\bibfnamefont{R.}~\bibnamefont{Verzicco}},\ and\ \bibinfo {author}
  {\bibfnamefont{D.}~\bibnamefont{Lohse}},\ }%
  \bibfield{title}{%
  \enquote{\bibinfo {title} {Optimal {{Taylor-Couette}} flow: direct numerical
  simulations},}\ }%
  \bibfield{journal}{%
  \bibinfo {journal} {J. Fluid Mech.}\ }%
  \textbf{\bibinfo {volume} {719}},\ \bibinfo {pages} {14--46} (\bibinfo {year}
  {2013})\BibitemShut{NoStop}%
\bibitem{ost14}%
  \BibitemOpen
  \bibfield{author}{%
  \bibinfo {author} {\bibfnamefont{R.}~\bibnamefont{Ostilla}}, \bibinfo
  {author} {\bibfnamefont{E.~P.}\ \bibnamefont{van~der Poel}}, \bibinfo
  {author} {\bibfnamefont{R.}~\bibnamefont{Verzicco}}, \bibinfo {author}
  {\bibfnamefont{S.}~\bibnamefont{Grossmann}},\ and\ \bibinfo {author}
  {\bibfnamefont{D.}~\bibnamefont{Lohse}},\ }%
  \bibfield{title}{%
  \enquote{\bibinfo {title} {Boundary layer dynamics at the transition between
  the classical and the ultimate regime of {{Taylor-Couette}} flow},}\ }%
  \bibfield{journal}{%
  \bibinfo {journal} {Phys. Fluids}\ }%
  \textbf{\bibinfo {volume} {x}},\ \bibinfo {pages} {y} (\bibinfo {year}
  {2014})\BibitemShut{NoStop}%
\bibitem{bra13}%
  \BibitemOpen
  \bibfield{author}{%
  \bibinfo {author} {\bibfnamefont{H.}~\bibnamefont{Brauckmann}}\ and\ \bibinfo
  {author} {\bibfnamefont{B.}~\bibnamefont{Eckhardt}},\ }%
  \bibfield{title}{%
  \enquote{\bibinfo {title} {{Direct Numerical Simulations of Local and Global
  Torque in Taylor-Couette Flow up to Re=30.000}},}\ }%
  \bibfield{journal}{%
  \bibinfo {journal} {J. Fluid Mech.}\ }%
  \textbf{\bibinfo {volume} {718}},\ \bibinfo {pages} {398--427} (\bibinfo
  {year} {2013})\BibitemShut{NoStop}%
\bibitem{Note1}%
  \BibitemOpen
  \bibinfo {note} {In the astrophysical context, e.g.\ ref.\ \cite {dub05},
  often an extra factor $2/(1+\eta )$ is used in this
  definition.}\BibitemShut{Stop}%
\bibitem{tay17}%
  \BibitemOpen
  \bibfield{author}{%
  \bibinfo {author} {\bibfnamefont{G.~I.}\ \bibnamefont{Taylor}},\ }%
  \bibfield{title}{%
  \enquote{\bibinfo {title} {Motion of solids in fluids when the flow is not
  irrotational},}\ }%
  \bibfield{journal}{%
  \bibinfo {journal} {Proc. R. Soc. Lond. A}\ }%
  \textbf{\bibinfo {volume} {93}},\ \bibinfo {pages} {92--113} (\bibinfo {year}
  {1917})\BibitemShut{NoStop}%
\bibitem{pro16}%
  \BibitemOpen
  \bibfield{author}{%
  \bibinfo {author} {\bibfnamefont{J.}~\bibnamefont{Proudman}},\ }%
  \bibfield{title}{%
  \enquote{\bibinfo {title} {On the motion of solids in a liquid possessing
  vorticity},}\ }%
  \bibfield{journal}{%
  \bibinfo {journal} {Proc. R. Soc. Lond. A}\ }%
  \textbf{\bibinfo {volume} {92}},\ \bibinfo {pages} {408--424} (\bibinfo
  {year} {1916})\BibitemShut{NoStop}%
\bibitem{cha81}%
  \BibitemOpen
  \bibfield{author}{%
  \bibinfo {author} {\bibfnamefont{S.}~\bibnamefont{Chandrasekhar}},\ }%
  \emph{\bibinfo {title} {Hydrodynamic and Hydromagnetic Stability}}\ (\bibinfo
  {publisher} {Dover},\ \bibinfo {address} {New York},\ \bibinfo {year}
  {1981})\BibitemShut{NoStop}%
\bibitem{eck07b}%
  \BibitemOpen
  \bibfield{author}{%
  \bibinfo {author} {\bibfnamefont{B.}~\bibnamefont{Eckhardt}}, \bibinfo
  {author} {\bibfnamefont{S.}~\bibnamefont{Grossmann}},\ and\ \bibinfo {author}
  {\bibfnamefont{D.}~\bibnamefont{Lohse}},\ }%
  \bibfield{title}{%
  \enquote{\bibinfo {title} {Torque scaling in turbulent taylor-couette flow
  between independently rotating cylinders},}\ }%
  \bibfield{journal}{%
  \bibinfo {journal} {J. Fluid Mech.}\ }%
  \textbf{\bibinfo {volume} {581}},\ \bibinfo {pages} {221--250} (\bibinfo
  {year} {2007})\BibitemShut{NoStop}%
\bibitem{gro00}%
  \BibitemOpen
  \bibfield{author}{%
  \bibinfo {author} {\bibfnamefont{S.}~\bibnamefont{Grossmann}}\ and\ \bibinfo
  {author} {\bibfnamefont{D.}~\bibnamefont{Lohse}},\ }%
  \bibfield{title}{%
  \enquote{\bibinfo {title} {Scaling in thermal convection: A unifying view},}\
  }%
  \bibfield{journal}{%
  \bibinfo {journal} {J. Fluid. Mech.}\ }%
  \textbf{\bibinfo {volume} {407}},\ \bibinfo {pages} {27--56} (\bibinfo {year}
  {2000})\BibitemShut{NoStop}%
\bibitem{gro01}%
  \BibitemOpen
  \bibfield{author}{%
  \bibinfo {author} {\bibfnamefont{S.}~\bibnamefont{Grossmann}}\ and\ \bibinfo
  {author} {\bibfnamefont{D.}~\bibnamefont{Lohse}},\ }%
  \bibfield{title}{%
  \enquote{\bibinfo {title} {Thermal convection for large {{Prandtl}}
  number},}\ }%
  \bibfield{journal}{%
  \bibinfo {journal} {Phys. Rev. Lett.}\ }%
  \textbf{\bibinfo {volume} {86}},\ \bibinfo {pages} {3316--3319} (\bibinfo
  {year} {2001})\BibitemShut{NoStop}%
\bibitem{Note2}%
  \BibitemOpen
  \bibinfo {note} {I.e., ignoring possible crossovers to an ultimate regime in
  which the energy dissipation rate may be dominated by the bulk dissipation
  \cite {gro00,gro11} $\epsilon _{bulk}\sim \nu ^3 d^{-4} Re_s^3$. In that
  regime $\protect \mathaccentV {tilde}07E\tau $ would become independent on
  $Re_s$.}\BibitemShut{Stop}%
\bibitem{wit74}%
  \BibitemOpen
  \bibfield{author}{%
  \bibinfo {author} {\bibfnamefont{E.~M.}\ \bibnamefont{Withjack}}\ and\
  \bibinfo {author} {\bibfnamefont{C.~F.}\ \bibnamefont{Chen}},\ }%
  \bibfield{title}{%
  \enquote{\bibinfo {title} {An experimental study of couette instability of
  stratified fluids},}\ }%
  \bibfield{journal}{%
  \bibinfo {journal} {J. Fluid Mech.}\ }%
  \textbf{\bibinfo {volume} {66}},\ \bibinfo {pages} {725--737} (\bibinfo
  {year} {1974})\BibitemShut{NoStop}%
\bibitem{kla03}%
  \BibitemOpen
  \bibfield{author}{%
  \bibinfo {author} {\bibfnamefont{H.H.}\ \bibnamefont{Klahr}}\ and\ \bibinfo
  {author} {\bibfnamefont{P.}~\bibnamefont{Bodenheimer}},\ }%
  \bibfield{title}{%
  \enquote{\bibinfo {title} {Turbulence in accretion disks: Vorticity
  generation and angular momentum transport via the global baroclinic
  instability},}\ }%
  \bibfield{journal}{%
  \bibinfo {journal} {The Astrophysical Journal}\ }%
  \textbf{\bibinfo {volume} {508}},\ \bibinfo {pages} {869} (\bibinfo {year}
  {2003})\BibitemShut{NoStop}%
\bibitem{pac81}%
  \BibitemOpen
  \bibfield{author}{%
  \bibinfo {author} {\bibfnamefont{B.}~\bibnamefont{Paczynski}}\ and\ \bibinfo
  {author} {\bibfnamefont{G.}~\bibnamefont{Bisnovatyi-Kogan}},\ }%
  \bibfield{title}{%
  \enquote{\bibinfo {title} {A model of a thin accretion disk around a black
  hole},}\ }%
  \bibfield{journal}{%
  \bibinfo {journal} {Acta Astronomica}\ }%
  \textbf{\bibinfo {volume} {31}} (\bibinfo {year} {1981})\BibitemShut{NoStop}%
\bibitem{gro11}%
  \BibitemOpen
  \bibfield{author}{%
  \bibinfo {author} {\bibfnamefont{S.}~\bibnamefont{Grossmann}}\ and\ \bibinfo
  {author} {\bibfnamefont{D.}~\bibnamefont{Lohse}},\ }%
  \bibfield{title}{%
  \enquote{\bibinfo {title} {Multiple scaling in the ultimate regime of thermal
  convection},}\ }%
  \bibfield{journal}{%
  \bibinfo {journal} {Phys. Fluids}\ }%
  \textbf{\bibinfo {volume} {23}},\ \bibinfo {pages} {045108} (\bibinfo {year}
  {2011})\BibitemShut{NoStop}%
\end{thebibliography}%

\end{document}